\documentclass[a4paper,USenglish,cleveref, autoref]{lipics-v2019}

\usepackage{url}

\usepackage{breakurl}
\PassOptionsToPackage{breaklinks}{hyperref}

\title{From Theory to Systems: A Grounded Approach to Programming Language Education} 

\titlerunning{From Theory to Systems}

\author{Will Crichton}{Stanford University}{wcrichto@cs.stanford.edu}{}{}

\authorrunning{W. Crichton}

\Copyright{Will Crichton}



\category{}

\relatedversion{}

\supplement{}

\nolinenumbers 

\hideLIPIcs  

\EventEditors{John Q. Open and Joan R. Access}
\EventNoEds{2}
\EventLongTitle{42nd Conference on Very Important Topics (CVIT 2016)}
\EventShortTitle{CVIT 2016}
\EventAcronym{CVIT}
\EventYear{2016}
\EventDate{December 24--27, 2016}
\EventLocation{Little Whinging, United Kingdom}
\EventLogo{}
\SeriesVolume{42}
\ArticleNo{23}

\begin{document}

\maketitle

\begin{abstract}
I present a new approach to teaching a graduate-level programming languages course focused on using systems programming ideas and languages like WebAssembly and Rust to motivate PL theory. Drawing on students' prior experience with low-level languages, the course shows how type systems and PL theory are used to avoid tricky real-world errors that students encounter in practice. I reflect on the curricular design and lessons learned from two years of teaching at Stanford, showing that integrating systems ideas can provide students a more grounded and enjoyable education in programming languages. The curriculum, course notes, and assignments are freely available: \url{http://cs242.stanford.edu/f18/}
\end{abstract}

\section{Introduction}

A deep divide exists today between courses on programming languages across academia. Some PL courses focus on theory and formal methods, such as CMU\,\cite{cmusite}, UPenn\,\cite{pennsite}, and Princeton\,\cite{princetonsite}. Some combine PL and compilers to cover the implementation of grammars, typecheckers, and interpreters like Northeastern\,\cite{northeasternsite}, UC Berkeley\,\cite{berkeleysite}, and UIUC\,\cite{uiucsite}. Others yet focus on teaching existing paradigms of programming languages like functional and logic programming at places like UW\,\cite{uwsite} and (formerly) Stanford\,\cite{stanfordsite}. (And others still blaze new trails, like Brown\,\cite{pombrio2017teaching}.)

While I strongly believe in the educational value of programming language theory, I worry that the theory-oriented PL courses (e.g. courses using \cite{pierce2002types, harper2016practical}) end up only interesting the math-oriented programmers that appreciate the intrinsic beauty of concise programming models and proofs. Anecdotally, my peers and I experienced this during undergrad at CMU. I heard sentiments about PL theory like, ``I get it, it's cool, but I'm not designing functional programming languages or doing PL research, so it's not for me.''

By contrast, while implementation-oriented or ``programming language zoo'' courses are  looked down upon by theorists, these courses are undeniably \textit{fun} in ways theory courses often are not. The journey of discovering new languages, the excitement of exploring unfamiliar paradigms---for me, these moments, arcs, and emotions carried my passion for programming languages through the years moreso than abstract theory. Motivating students to continue learning and using what we teach is just as important as the content itself.

Over the last two years, I redesigned Stanford's graduate PL course, CS 242, in an attempt to bridge this divide. I developed a novel curriculum that integrates modern systems programming languages like WebAssembly and Rust to contextualize and motivate discussions of functional programming, type theory, and operational semantics. In this paper, I will discuss the key design decisions of this curriculum and my experience teaching it to 70 Stanford CS students.

\section{Pedagogy}

The target audience for this course is the average Stanford senior undergraduate or master's student. This student has completed the computer science core (data structures, computer systems, algorithms, discrete mathematics, etc.) and has no formal exposure to functional programming (CS 242 is the only Stanford course that teaches FP). They probably intend to become a software engineer at a large tech company after graduation, or perhaps already are---Google and Facebook engineers have taken CS 242 through job training programs. This student does not necessarily come predisposed with a love of mathematical theory, of functional programming, or of programming languages at all. 

For such a student, I believe the primary benefit of learning PL theory is to provide a simple model of fundamental concepts in computation. I call this ``computational literacy,'' i.e.\ not just knowing how to program, but understanding the essence of ideas like variables, functions, and types. From simplicity comes two key benefits: 

    \noindent\textbf{1. Improving mental models of computation}: most students learn programming ad hoc through courses, extracurriculars, and internships. These experiences provide a working knowledge of computational concepts through the lens of existing programming tools, but such concepts often carry significant baggage due to reasons of poor design, legacy code, or limitations of the underlying hardware. For example, a student who learns C assumes a mental model of functions that involves multiple/variadic arguments, void returns, top-level vs. nested functions, recursion, control flow through early returns, the call stack, and so on. By contrast, a student who learns the lambda calculus can understand functions solely as abstraction of code over a single variable (along with variable scope). I strongly believe that providing students a clear mental model of computational concepts should be a core mission of not just a PL course, but of a broader CS curriculum. I suspect that these concepts are essential in acquisition and transfer of computational skills like building new algorithms or learning new programming languages, although our current understanding of the programmer's psyche is perilously insufficient to fully justify this claim. 
    
    \noindent\textbf{2. Enabling formal reasoning}: when a computational model is simple (has few rules, syntactic constructs, edge cases, etc.), then the model becomes significantly easier to analyze for correctness, computational complexity, and many other such properties. For students both using and potentially designing computational models, simplicity becomes a criterion for understanding and evaluating the amenability of a model to analysis. Moreover, students can gain exposure to a wider variety of program analysis techniques than just those traditionally applied to programming languages. Existential types, ownership types, session types, refinement types, etc. all serve to expand a student's idea of the realm of possibilities in language design. 

My goal for CS 242 is to demonstrate how programming language theory provides a simpler view on computational concepts, and to contextualize the benefits of simplicity against problems familiar to students. All CS students at Stanford are required to take an introduction to computer systems course, i.e.\ assembly, C, pointers and memory, bit manipulation, and so on. This provides an opportunity to ground the theoretical framework of PL in their concrete experiences writing low-level programs, showing that PL theory could solve problems they empathize with. What if a static type system could prevent you from ever getting a segfault? What if assembly actually had a type system? Answering these questions both provides concrete applications of PL theory in practice, but also motivates students by addressing problems they understand viscerally: just ask any student how long they've spent staring at x86 or debugging a segfault.

\section{Curriculum}

CS 242 has seven arcs across eight weeks of core content, roughly 50/50 between theory and systems. The theory arcs are largely derived from a subset of \textit{Types and Programming Languages}\,\cite{pierce2002types}, while the systems arcs are entirely my own invention. Students are expected to do 12-15 hours of work per week including lectures and weekly assignments.

\subsection{Logic and semantics}

We first establish the logical system of judgments and inference rules, followed by a presentation of the untyped lambda calculus and its operational semantics. The assignment, like many in this course, is structured around introducing new concepts and getting students to transfer knowledge from lecture to new contexts. This is as opposed to, say, just implementing an interpreter for the abstractions defined in class. Rather than teach students the Church encoding in class, we provide an untyped lambda calculus interpreter with OCaml-ish syntax and have students build the encoding themselves. This requires students to transfer an understanding of the lambda calculus execution model from the simple examples shown in class to the more abstract ones in the Church encoding. 

Similarly, to test the student's ability to interpret semantic rules, we introduce the semantics for a dynamically-scoped lambda calculus and ask students to complete a proof of an execution trace and provide a semantic rule for a new syntactic construct. This both prepares students to understand proof contexts used later for typing judgments, and it also demonstrates how to reformulate language semantics under a different design decision, i.e. dynamic vs. lexical scoping. 
    
\subsection{Type theory}

We next move to the basics of static semantics and the typed lambda calculus. We discuss structural induction along with progress and preservation, both to highlight a topic of intrinsic interest (demonstrating the safety of a language), but also to show a first example of proving a complex property about a language. Given the limited course time, we do not discuss proofs of totality or other such properties. Because functional programming is not a course prerequisite, we simultaneously introduce OCaml, framing the language as a featureful typed lambda calculus interpreter. 

For the assignment, students implement an interpreter for the typed lambda calculus with algebraic data types in OCaml, learning both languages at the same time. This more standard exercise is supplemented with a language design challenge: the student is given the static and dynamic semantics for two proposed language extensions (let bindings and induction over naturals). One violates progress or preservation and the other does not, and the student must figure out which is which and provide the corresponding proof/counterexample. From my interactions with students, this kind of written exercise is a necessary complement to the interpreter, since it's very possible for students to successfully write the OCaml program without fully internalizing how the lambda calculus actually works.

\subsection{Functional programming} 

We then discuss three important areas in type systems: algebraic data types, recursion, and parametric polymorphism. On each topic, I give a practical introduction as one might find in an introduction to functional programming course, followed by a discussion of formal semantics in the typed lambda calculus. For example, when teaching ADTs, I will give examples of how sum types in OCaml better represent error conditions than error codes and null pointers, followed by a presentation of their static and dynamic semantics in the typed lambda calculus along with a progress/preservation proof. 

For this assignment, students implement a polymorphic collections library in OCaml and translate their implementation by hand into typed lambda calculus. The coding portion is of the kind one would find in an introduction to functional programming class, the goal being to have students write functional programs of moderate complexity that aren't interpreters. The written portion is part of a broader  strategy of learning by analogy, i.e.\ students are more likely to recognize the similarities of the OCaml and lambda calculus representations if asked to explicitly translate between the two. Similarly in lecture, if I show a new functional programming construct, I will consistently translate back to a familiar language like C, Java, or Python to help students relate to languages and problems they know. 
    
\subsection{WebAssembly} 

This is the major divergence point with existing curricula. We start our shift into systems by asking: now that we understand the basics of a mathematical framework for programming languages, how can we use this knowledge to improve the design of real-world languages? WebAssembly is a perfect example, as it is a popular, cutting-edge language for high performance web development, and its authors have presented a formal semantics\,\cite{haas2017bringing}  and a proof of progress and preservation\,\cite{watt2018mechanising} for the language. WebAssembly introduces the notion that even assembly languages don't have to be unsafe, and could support a type system with careful language design (e.g.\ no jumps). I use its semantics as a motivating example to introduce formal representations of mutability and non-local control flow. The actual semantics are too complicated to be feasibly taught in a single lecture, so I instead designed a distilled version of the language, e.g. separating the instruction sequence from the value stack and eliminating block contexts for branch semantics\,\cite{wasm_semantics}.

The assignment has students implement a memory allocator using plain WebAssembly and formulate semantics for adding exceptions into the language. The allocator uses a simple implicit free list that all students are familiar with from their introduction to systems course, but this time recast into WebAssembly instead of C. This helps students internalize WebAssembly semantics and identify the key differences with x86 and C like structured branching. The written portion asks students to provide operational semantics for a basic exception mechanism given an English specification. This task both provides students an opportunity to design semantics for a nontrivial language feature, and also highlights the difference between static/dynamic control flow with WebAssembly's branches versus the proposed exceptions.
    
\subsection{Rust} 

While we use WebAssembly to demonstrate a direct application of PL theory, we introduce Rust to demonstrate more broadly the utility of type systems in low-level programming. We focus primarily on using the borrow checker (i.e. affine and ownership types) for resource management, especially with respect to memory safety and concurrency. I also emphasize the trait system for understanding a more compositional approach to modular programming than the traditional inheritance-based object-oriented style. I do not use formal semantics to teach either topic, as that would take up too much time. My goal is just to convince students that tricky systems problems like avoiding segfaults and data races can be categorically eliminated through static analysis combined with careful language design. 

This arc covers two assignments: first, students implement a WebAssembly interpreter in Rust to learn Rust basics and reinforce knowledge of WebAssembly semantics. The WebAssembly semantics are significantly more complex than the typed lambda calculus, having local mutable variables, a function call stack, and an addressable memory. Students experience the contrast between implementing a simple interpreter in OCaml versus a complex interpreter in Rust. 

Second, students implement a combinator-based futures library\,\cite{futures} to explore the interaction of memory management, concurrency, and traits in a DSL. Most systems concepts like concurrency and memory management are taught in languages with impoverished type systems like C, which shapes the student's mental model around ugly interfaces like pthreads and epoll. This assignment shows how a proper type system provides the foundation for expressing domain concepts with better abstractions, e.g. using parametric polymorphism instead of \verb|void*| for generic thread callbacks, or using sum types and pattern matching to manage communication between threads instead of tagged unions or \verb|#define|. 
    
\subsection{Session types} In the same way UW's ``Hack Your Language'' course\,\cite{hacklangsite} promotes an ethos where anyone can implement an embedded DSL, I want to convey the same lesson but for domain-specific type theories. To do this by example, we introduce two-party session types\,\cite{takeuchi1994interaction} along with their standard theory (e.g. computing duals), and then walk through an implementation of session types in Rust\,\cite{jespersen2015session}. Session types pull together many of the course's lessons: articulating a type theory using mathematical notation, translating inference rules into code with traits, avoiding dangerous aliasing with affine types, implementing lexically scoped variables with de Bruijn indices, and defining recursion through fix points. We also discuss more general notions of typestate, such as statically verified finite state machines. 

Students then design and build a session-typed (simplified) TCP implementation in Rust. We provide a high-level English specification and an example communication trace with sufficient detail to ensure that, although the students design the session type themselves, there is only one possible type they can correctly derive. The student's implementation contains the session type and an implementation of the server and client, which we test by compiling an adversarial server/client against their code, checking that the session types match, and then verifying the correctness properties not captured in the type.
    
\subsection{Dynamic typing}

In the course's final arc, I conclude by asking: static types are great, but what do we lose by having them in our language? We review many of the language constructs in the course, e.g. polymorphic types, exceptions, etc., and consider how dynamically-typed languages allow programmers to get these features ``for free'' while statically-typed languages require significant machinery in the compiler. I show how dynamically-typed languages make it easy to implement new language features. We specifically look at building a multiple-inheritance object-oriented programming system from scratch within Lua using prototypes. 

For their final two-week-long assignment, students implement a variant of this OOP system, build a small ASCII-art adventure game on top of it, then rewrite core components in Rust using Lua's C API. In the first section, students implement a class-based (as opposed to prototype-based) OOP system so students can draw parallels to the litany of class-based OOP languages they know. In the second section, we provide students most of the infrastructure for running the game, except that it relies on their class library (thus providing a substantial integration test). We have students implement a simple AI by introducing coroutines to represent state machines. Finally, the game's vector library is rewritten in Rust to show an example of how Lua's API enables the integration of statically typed, GC-free code with dynamically typed, garbage-collected code.

\subsection{Final project}

CS 242 does not have a midterm---the majority of its grade comes from the eight assignments described above. I am still in the process of determining the the best way to do a final. In the first iteration, students did final projects for three weeks. Students enjoyed the opportunity to do a deep dive on a PL-related subject of personal interest, and two-thirds said they preferred doing a project over an exam. However, the prompt was too open-ended and the three weeks required cut too much into class time that would otherwise be used for more directed assignments.

On the latest iteration, I had students learn to use the Lean theorem prover\,\cite{de2015lean} on their own and prove a number of simple theorems (i.e. a take-home final, still not an exam). The goal of the final was to show that learning about PL theory and functional programming would make students more productive in settings of self-directed learning that involved foreign PL concepts, such as grappling with dependent types and the Curry-Howard correspondence. While students were able to effectively pick up the core concepts just by reading the Lean documentation (a reflection perhaps on the high quality of the Lean documentation!), several students provided feedback that the final felt disjointed from the course.  In future iterations, I plan to make the final more systems-related while still preserving the same themes, perhaps by learning and using TLA+ to verify an actual distributed system instead of arbitrary logic theorems.

\section{Outcomes}

I taught this version of the curriculum in Autumn 2018 to a class of 77 Stanford students, mostly computer science majors, and mostly masters and upper-level undergraduate students. 85\% of the students provided anonymized course feedback, mostly numeric with a subset of students leaving comments. 

Students enjoyed the course and felt they learned a good amount of material. Students were asked ``How much did you learn from this course?'' with responses of ``A great deal'' (5), ``A lot'' (4), ``A moderate amount'' (3), ``A little'' (2), ``Nothing'' (1). The response average was 4.3 with 87\% of students answering 4 or above. When prompted to articulate specific skills learned, most respondents wrote about: 1) PL theory and proofs about programs, 2) functional programming and new models of computation, or 3) learning to quickly acquire new languages. I believe this variety of responses reflects positively on the curriculum---different students enter with different personal learning goals, and the course's breadth allows students to find the aspect of programming languages they enjoy most.

The survey also prompted students with ``What would you like to say about this course to a student who is considering taking it in the future?'' I reproduce a few responses below.
\begin{quote}
    [CS 242] reflects on the translation that occurs between the model of computation you have in your head and how you express that model of computation using the tools a programming language gives you. \\
    
    It's still not really clear to me how the Lambda Calculus practically relates to the rest of the course, so the first few weeks were a bit frustrating, but I found it more enjoyable after that. \\

    It was not what I expected and was a lot of work (office hours were a blessing) but it introduced me (thoroughly!) to functional programming concepts that I am now looking into further on my own because I am interested in different ways of thinking about computation and how that will affect the way I approach programming problems in the future. \\
    
    One of my favorite classes at Stanford (my 4th year for context) and has really opened my eyes to the world of PL and how I actually really enjoy it and might be looking into pursuing research or further education in PL related work. \\

    This is one of the best CS classes I've taken at Stanford--if not hands down the best. Lectures are always so exciting, and the homework is a ton of fun. I really like the mix of theory and systems-level stuff, as well as the mix of programming and written assignments, where both feel like they are valuable and adding to the other. I also feel like I'm being challenged to think in different ways than I do in most CS classes (mostly because of the functional paradigm). 
\end{quote}

\noindent I firmly believe that students learn most from assignments. Many students enjoyed the course likely because I put significant effort into developing assignments that didn't just reinforce the course content, but did so in an exciting way. This is my biggest critique of the definitional interpreters approach to PL education: while it showcases the content, it gets boring quickly. Implementing TCP, an adventure game, a memory allocator---this variety keeps students engaged while they pick up new PL concepts. Moreover, the focus on real-world systems promotes creativity in assignments, as you naturally get access to a wider range of libraries and applications than when using niche teaching languages.

\section{Discussion}

Overall, I consider the focus on bridging theory and systems to be a major success. Students leave the course with different takeaways on the most important material, but almost everyone leaves understanding something about the role and importance of programming language theory in practice. Going forward, I will continue working to draw explicit connections between the different parts of the course---as one comment mentioned, some students are still struggling to see the relationships between the abstract ideas of the lambda calculus and the everyday concerns of systems programming.

In particular, I would like to refine the lecture and assignments to reflect the role of PL theory in the process of designing computational systems. Operational semantics and the like shouldn't just be seen as verification tools for functional-esque programming languages, but instead as a design tool for any system with a compositional model. Proofs like progress and preservation aren't check boxes on the way to language safety, but part of reasoning about unforeseen edge cases. Put another way: I believe we should seek to integrate lessons from the theorem-prover-oriented PL courses (e.g. Princeton\,\cite{princetonsite}, UPenn\,\cite{pennsite}) where possible.

Lastly, if any instructors would like to use part or all of this curriculum in their own courses, I have made all course materials are freely available. The course website (\url{http://cs242.stanford.edu/f18/}) contains a mini-textbook of typeset notes for each lecture. Starter code for all assignments is open-source on GitHub (\url{https://github.com/stanford-cs242/f18-assignments}). \\


\noindent \textbf{Acknowledgments.} I am eternally grateful to my advisor, Pat Hanrahan, for his support and feedback through the years. His positivity, encouragement, and wisdom continue to push me and this course to ever greater heights.



\bibliography{bib}

\end{document}